# Observation of current-induced switching in non-collinear antiferromagnetic IrMn$_3$ by differential voltage measurements

Sevdenur Arpaci [1,2,†], Victor Lopez-Dominguez [1†*], Jiacheng Shi [1], Luis Sánchez-Tejerina [3], Francesca Garesci [4], Chulin Wang [1], Xueting Yan [1], Vinod K. Sangwan [5], Matthew Grayson [1,2], Mark C. Hersam [1,2,5,6], Giovanni Finocchio [3*], Pedram Khalili Amiri [1,2*]

[1] Department of Electrical and Computer Engineering, Northwestern University, Evanston, Illinois 60208, United States of America

[2] Graduate Program in Applied Physics, Northwestern University, Evanston, Illinois 60208, United States of America

[3] Department of Mathematical and Computer Sciences, Physical Sciences and Earth Sciences, University of Messina, Messina 98166, Italy

[4] Department of Engineering, University of Messina, Messina 98166, Italy

[5] Department of Materials Science and Engineering, Northwestern University, Evanston, Illinois 60208, United States of America

[6] Department of Chemistry, Northwestern University, Evanston, Illinois 60208, United States of America

[†] These authors contributed equally to this work.

[*] Correspondence and requests for materials should be addressed to V.L.-D., G.F., or P.K.A.

[*] Email: victor@northwestern.edu , gfinocchio@unime.it , pedram@northwestern.edu

**Abstract**

There is accelerating interest in developing memory devices using antiferromagnetic (AFM) materials, motivated by the possibility for electrically controlling AFM order via spin-orbit torques, and its read-out via magnetoresistive effects. Recent studies have shown, however, that high current densities create non-magnetic contributions to resistive switching signals in AFM/heavy metal (AFM/HM) bilayers, complicating their interpretation. Here we introduce an experimental protocol to unambiguously distinguish current-induced magnetic and nonmagnetic switching signals in AFM/HM structures, and demonstrate it in $IrMn_3$/Pt devices. A six-terminal double-cross device is constructed, with an $IrMn_3$ pillar placed on one cross. The differential voltage is measured between the two crosses with and without $IrMn_3$ after each switching attempt. For a wide range of current densities, reversible switching is observed only when write currents pass through the cross with the $IrMn_3$ pillar, eliminating any possibility of non-magnetic switching artifacts. Micromagnetic simulations support our findings, indicating a complex domain-mediated switching process.

**Introduction**

Antiferromagnetic materials (AFMs) provide a pathway to overcome the limitations of ferromagnet (FM)-based spintronic devices, thereby enabling new applications of spintronics in memory, computing, and terahertz electronics [1-6]. Their robustness against external magnetic fields, potential for high-density data storage, absence of stray fields, ultrafast dynamics, and high energy efficiency make them an excellent candidate for ultrafast magnetic random access memory [7, 8], particularly at deeply scaled technology nodes that are important to emerging applications in artificial intelligence and unconventional (non-von Neumann) computing [9]. At the same time, their large spin coherence length [1, 5, 10, 11] and excellent magnon propagation [12-15] make antiferromagnets promising candidates for information transmission in emerging computing concepts based on spintronics. Beyond computing applications, the ultrafast exchange-dominated dynamics of antiferromagnets also make them potentially of interest for the realization of room-temperature, electrically tunable, and narrowband terahertz sources and detectors [3, 16-18].

Making memory units with AFMs requires experimental techniques to: (i) Electrically manipulate the AFM order (Néel vector) in a silicon-compatible material (i.e., *write* information), as well as to (ii) Electrically read out the orientation of the Néel vector (i.e., *read* information) once it has been modified. The first requirement was initially demonstrated in antiferromagnetic films with broken inversion symmetry (CuMnAs [8, 19-21], Mn$_2$Au [22-26]) where a damping-like Néel spin-orbit torque (SOT) was generated due to the current flowing in the bulk of the material, resulting in the motion of AFM domains. This was followed by studies in AFM insulators (NiO [27-31], Fe$_2$O$_3$ [32], CoO [33]), as well as thin films [34, 35] and pillars [36] of metallic AFM materials interfaced with a heavy metal (HM) with large spin-orbit coupling (typically, Pt), where the mechanism of AFM order manipulation was the interfacial spin-orbit torque from the HM. In both types of experiments, the manipulation of AFM domains was directly observed using x-ray magnetic linear dichroism (XMLD) [20, 37] and spin Seebeck microscopy [38] measurements. However, while helping to reveal the physics of AFM switching by electric current, these techniques are not directly applicable to practical AFM memory devices, where an electrical readout mechanism is needed. This second requirement – reliable electrical readout of the AFM order using magnetoresistive signals (i.e. the dependence of electrical resistance on the Néel vector orientation) – has proven to be challenging, and is the focus of the present work.

Previous electrical switching experiments in AFM/HM systems have been performed on a single-cross geometry with the HM as the bottom layer. However, several recent reports have indicated that some of the electrically measured switching signals in such test structures may have non-magnetic origins due to thermal effects and atomic motion (e.g. electromigration) in the HM layer, particularly in cases where the applied current densities were high [39-41]. Chiang et al. showed that these non-magnetic resistive switching signals increased as the heat conductivity of the substrate reduced, pointing to the role of heat gradients in facilitating electromigration in Pt, which in turn gives rise to a switching signature that is not reflective of AFM switching [39]. A recent work by Cheng et al. [32] focusing on the Pt/Fe$_2$O$_3$ system showed that while switching at intermediate current densities may indeed be of magnetic origin, switching at higher current densities was dominated by signals from the Pt layer. Notably, in this material system, the antiferromagnetic switching was accompanied by a square-shaped behavior where the readout voltage was independent of the number of pulses applied in a particular direction, whereas the non-magnetic switching of the Pt exhibited a sawtooth behavior, where increasing the number of write pulses in a particular direction gradually increased or decreased the magnitude of the readout voltage. It is noteworthy that a similar square-shaped switching signature was also observed due to AFM order manipulation in the PtMn/Pt and PtMn/Ta systems [36]. Nonetheless, both for device applications and fundamental studies of AFM switching, it remains a challenge to reliably and directly separate the nonmagnetic and magnetic contributions to resistive switching signals in AFM/HM structures. This is particularly difficult for AFM/HM devices because, unlike FM/HM structures, magnetoresistance effects are typically smaller in AFMs, switching current densities are higher, and there is typically no possibility to manipulate (or saturate) the magnetic order with magnetic fields, something which can be routinely done with modest magnetic fields for FM/HM samples.

Here we present a new experimental protocol based on differential measurements, which allows one to systematically and robustly separate the switching signals originating from the AFM and the HM layers in AFM/HM material systems, thereby demonstrating the magnetic origin of the switching. The antiferromagnetic material that we choose for this demonstration is IrMn$_3$, grown on Pt as the HM, with both layers having a predominant (111) texture. This choice is in part motivated by the fact that this metallic AFM is compatible with conventional semiconductor manufacturing techniques, and can be deposited on arbitrary substrates (in our case, thermally oxidized silicon) in poly-crystalline form using ultra-high vacuum sputter deposition. At the same time, IrMn$_3$ is a non-collinear antiferromagnet with a range of intriguing transport characteristics, including a large spin Hall conductivity [42, 43], a large anomalous Hall effect in the absence of macroscopic magnetization [44-47], a theoretically predicted (ferromagnetic-like) spin-polarized charge current [48], and has been used in devices exhibiting large tunneling anisotropic magnetoresistance (TAMR) [49], making it of interest for both memory (switching) and high-frequency devices based on antiferromagnets.

The test structure used for the differential measurements is shown in Fig. 1. It consists of a double-cross Pt structure with six terminals, where an IrMn$_3$ pillar is placed on one of the two crosses. This configuration allows for revealing the difference between signals coming from the bilayer IrMn$_3$/Pt structure and Pt. Importantly, we compare two crosses in close proximity within the same device, to minimize the effect of wafer-to-wafer and across-wafer variations that may arise if we compared Pt-only structures with IrMn$_3$/Pt devices from different wafers, or even when comparing devices from different parts of the same wafer. By using this test structure, we demonstrate switching in IrMn$_3$ pillars with diameters of 6 µm and 4 µm, using current pulses in opposite directions (180˚ switching) as well as using current pulses that are perpendicular to each other (90° switching). The results indicate that for a range of writing current densities (from 36 to 50 MA/cm$^2$) switching only occurs in the IrMn$_3$ pillar, with no appreciable signal coming from the adjacent Pt cross, while for higher current densities (~ 60 MA/cm$^2$), both crosses show switching, indicating a signal that originates in the Pt layer. Interestingly, we also observe that the switching in the antiferromagnetic IrMn$_3$ pillars [42, 43] in fact has a sawtooth behavior, meaning that the shape of the readout signal by itself is not a reliable indicator of whether or not the switching is magnetic in origin. We further study the dependence of the switching behavior on the writing pulse number, which indicates a thermally activated domain wall motion mechanism underlying the observed switching. Micromagnetic simulations taking into account the cubic IrMn$_3$ anisotropy, and pinning potentials from the spatial distribution of the anisotropy, support our findings and qualitatively describe the switching processes arising from the manipulation of complex multidomain structures.

**Results**

*Materials and devices*

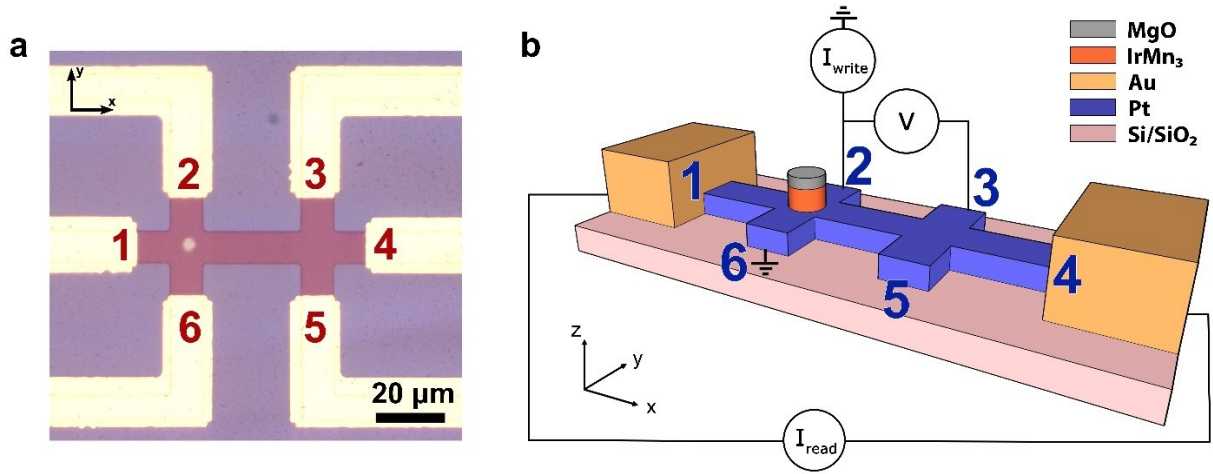

**Fig. 1. Design of double-cross device for differential voltage measurements. a** Microscope image of one of the studied devices. The device consists of two crosses with the same dimensions, where one has an IrMn$_3$ pillar and the second consists only of Pt. The electrodes are labeled from 1 to 6 to specify the directions in which the write current pulses are applied ($I^{i-j}_{write}$ refers to the write current applied between electrodes $i$ and $j$). In all cases, the electrodes 2 and 3 are used for measuring the differential readout voltage, $V^{2-3}_{read}$, while applying a small reading current, 100 µA, between electrodes 1 and 4. **b** Schematic representation of the device, showing the current sources ($I_{write}$ and $I_{read}$) and the nanovoltmeter ($V$) connections for one of the switching experiments ($I^{2-6}_{write}$). The other measurement configurations used in this study are outlined in the main text. The Cartesian coordinate system used here is also indicated as an inset.

The IrMn$_3$ film was grown by co-sputtering of high-purity Ir and Mn targets at different DC deposition powers of 10 W and 40 W, respectively, leading to a predominant IrMn$_3$ phase in the films with a (111) texture (See Supplementary Note 1 for the x-ray diffraction data). The AFM character of the resulting IrMn$_3$ layer was confirmed by characterizing the exchange bias in a thin Co layer adjacent to the IrMn$_3$. Further details on the deposition process are listed in the Methods section, while details of the exchange bias measurements are listed in Supplementary Note 2. The thickness of the IrMn$_3$ layer was determined by atomic force microscopy. After calibrating the deposition parameters, we constructed devices having a trilayer structure of Pt (5) / IrMn$_3$ (10) / MgO (2.5) (thickness in nanometers), grown on top of a thermally oxidized Si substrate. Here, the Pt layer acts as a SOT source to modify the AFM order of the IrMn$_3$ layer in response to electric current pulses, and the MgO is used as a protective capping layer.

A sketch of the patterned devices with the indication of the measurement scheme is shown in Fig. 1b. The device consists of two adjacent Pt crosses, connected through one of their four arms. The Pt layer is patterned by optical lithography in the form of the above-mentioned double-cross structure having 10 µm wide arms, and the IrMn$_3$ is patterned by optical lithography as a micrometer-scale pillar (with diameters of 4 or 6 µm), at the center of only one cross, as shown in Fig. 1a. All data shown in the main text are from a device with 6 µm IrMn$_3$ pillar diameter. Data for the 4 µm diameter case are consistent with these results, and are shown in Supplementary Note 3. The structure was completed with six Au pads to perform multidirectional electrical measurements at room temperature, labeled from 1 to 6 as shown in Figs. 1a-b.

*Experimental protocol*

The double-cross design allows for a differential DC resistance measurement between the two vertical arms (i.e. electrodes 2 and 3, hereafter referred to as $R_{2-3}$), when a small reading current (100 µA in our experiments) is applied along the long arm of the device (between electrodes 1 and 4, hereafter referred to as $I^{1-4}_{read}$). Since any effect associated with the Pt layer is expected to be the same in both arms, any variations of the voltage difference between the two vertical arms ($V^{2-3}_{read} = R_{2-3} I^{1-4}_{read}$) are expected to mainly depend on the resistance changes associated with the IrMn$_3$ pillar, in response to anisotropic magnetoresistance (AMR) and spin-Hall magnetoresistance (SMR) effects. Note that, apart from the new possibility for differential readout, this design can still be used in both of the usual configurations used for electrical switching (i.e. writing) of the AFM bit as will be shown in the paragraph below. We refer to these as 90° switching when the current pulses are applied perpendicular to each other, e.g. between contacts 1-4 and 2-6 or 3-5, and 180° switching when opposite write currents are sent through one of the vertical arms (contacts 2-6, as illustrated in Fig. 1b, or 3-5), or through the common horizontal arm (electrodes 1-4). Readout in all cases takes place through the same procedure, by reading $V^{2-3}_{read}$ while applying the reading current $I^{1-4}_{read}$.

In addition to differential measurements of the AFM switching in the IrMn$_3$ cross, this structure allows us to make in-situ reference measurements on the Pt cross within the same device. This is important since fabrication and patterning methods, even if nominally similar for two sets of samples, inevitably suffer from wafer-to-wafer and across-wafer variations in parameters such as the film thickness and lateral dimensions of metal lines, making it difficult to quantitatively compare control samples made on a different substrate to the AFM samples being studied. In the structure of Fig. 1, the control structure is already present in the final device and can be measured independently of the AFM arm to detect any signals originating from the Pt layer. As discussed in the next sub-section, this possibility is of particular interest for the 180° switching scheme when the writing pulse is only applied in the arm with the IrMn$_3$ pillar (electrodes 2 and 6 in Fig. 1) or without it (electrodes 3 and 5).

*Current-induced switching measurements*

We first show results for the 180° switching configuration in two cases: (i) When the write pulse is applied in opposite directions along the arm with the IrMn$_3$ pillar (between electrodes 2 and 6, $I^{2-6}_{write}$), i.e., the configuration shown in Fig. 1b, and (ii) a reference measurement when the write current is applied in the arm without the AFM element (between the electrodes 3 and 5, $I^{3-5}_{write}$). The amplitude of the pulses is 20 mA (corresponding to a current density of 40 MA/cm$^2$ in the Pt layer), and the pulse width is 1 ms. Each write attempt consists of applying two pulses in the same direction (e.g. from electrode 2 to 6, or 3 to 5), and then measuring the differential voltage between the two vertical arms ($V^{2-3}_{read} = R_{2-3} I^{1-4}_{read}$) using a reading current of 100 µA (for further details of the electrical measurement, see the Methods section). Next, this process is performed for write attempts with the reverse write current direction (e.g. from electrode 6 to 2, or 5 to 3 in the case of the Pt reference measurement).

The results of this 180° switching experiment are shown in Fig. 2a. The results clearly show a change of the readout voltage ($V^{2-3}_{read}$) when the write pulse is applied in the cross with the IrMn$_3$ pillar (data points indicated by squares), whereas no significant change in the readout voltage is observed when the write currents are applied in the arm containing only Pt (data points indicated by circles). This result unambiguously shows that the switching detected in the double-cross structure is only due to the IrMn$_3$ pillar, with an average readout voltage difference of ~ 8 µV. In addition, when the number of applied pulses in the same direction increases, the output voltage continuously changes, increasing when they are applied from electrode 2 to 6 (red points), and decreasing when the direction of the pulses is reversed (blue points). This gradual change indicates that the switching takes place by the reorientation of the domain structure in the IrMn$_3$ pillar. It is worth pointing out that this gradual switching behavior is similar to the previously detected sawtooth switching in various AFM material systems [19], while also bearing similarity to previously reported sawtooth-like non-magnetic signals in Pt-only devices [32, 39]. In the present experiment, however, the differential voltage measurement, combined with the in-situ repetition of the experiment in the Pt reference arm, allows us to conclude that the switching (for a write current of 20 mA, equal to 40 MA/cm$^2$) is exclusively due to the IrMn$_3$ pillar. It is worth noting that, therefore, the presence of a sawtooth signal shape by itself is not necessarily a reliable sign of a non-magnetic artefact (see also the micromagnetic calculations below for further discussion), and may indeed be associated with AFM switching, as is the case in our experiment.

Two additional differential voltage measurement methods are shown in Supplementary Notes 7-9, both of which provide results consistent with those presented here. The first method (Supplementary Note 7) uses a resistive voltage divider based on the Wheatstone bridge combined with an operational amplifier (OA), where the subtraction of two voltages from the two arms of the device (with and without IrMn$_3$) is amplified by the OA. The second measurement (Supplementary Notes 8 and 9) uses a different series of six-terminal devices, where two nominally identical IrMn$_3$ pillars are placed on both crosses of the device. The two pillars are switched by currents in opposite directions, and the overall output voltage corresponds to the subtraction of the individual voltage outputs from the two pillars.

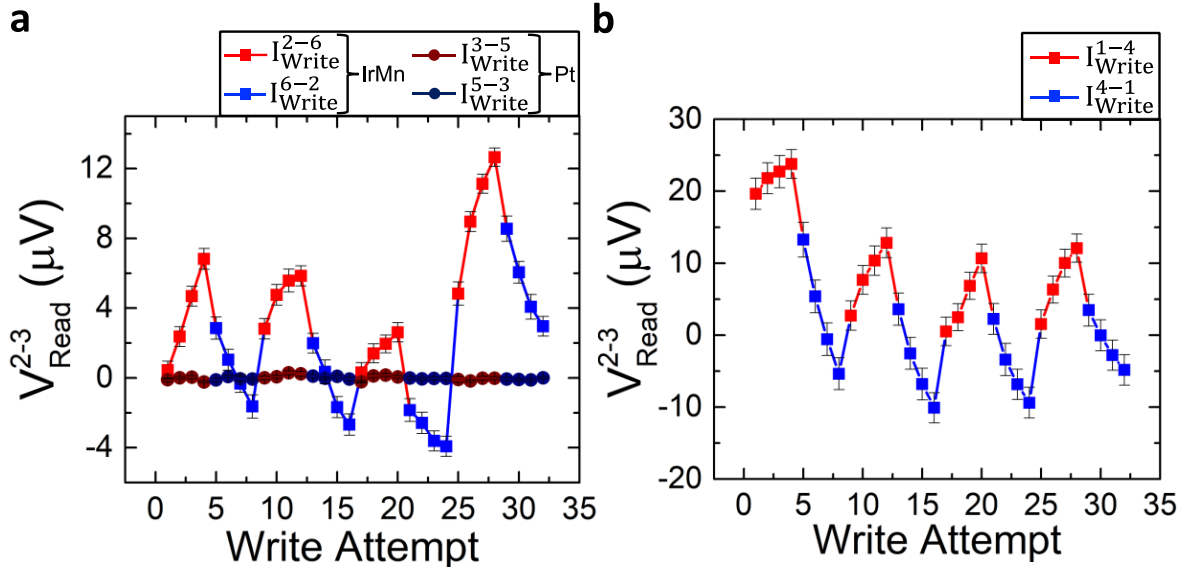

**Fig. 2. Measured differential voltage ($V^{2-3}_{Read}$) for the 180° switching configuration with 20 mA write pulses. a** The writing current is applied along the short arms of the double-cross structure. Square points refer to the case of a writing current applied in the arm with the IrMn$_3$ pillar (square red points from electrode 2 to 6, $I^{2-6}_{Write}$, and square blue points from electrode 6 to 2, $I^{6-2}_{Write}$), and the circle data points correspond to the arm only with Pt (circle brown points from electrode 3 to 5, $I^{3-5}_{write}$, and circle dark blue points from electrode 5 to 3, $I^{5-3}_{Write}$). The differential voltage of the device only shows a measurable switching signature when the current pulse is applied through the IrMn$_3$ pillar cross, i.e. in cases $I^{2-6}_{Write}$ and $I^{6-2}_{Write}$. **b** 180° switching when the writing current is applied from electrode 1 to 4 ($I^{1-4}_{write}$, square red points) or vice versa (square blue points). In this case, the write current pulse always passes through the section with the IrMn$_3$ pillar, showing a readout voltage with a similar switching behavior as the case of the write current $I^{2-6}_{write}$, but with larger voltage variation. Raw data for this figure (prior to subtraction of the background slope) can be found in Supplementary Note 3.

Next, we repeated the 180° switching experiment while applying both the writing and reading currents between the same horizontal electrodes (1 and 4 in Fig. 1). In this case, the writing pulse is applied through both of the crosses in the device, with and without the IrMn$_3$ pillar, hence allowing one to expect that switching would also be observed in this case. The results are shown in Fig. 2b, confirming this expectation. The results are qualitatively similar to the previous 180° switching experiment in the IrMn$_3$ arm, suggesting that the output voltage again changes because of the IrMn$_3$ pillar. The fact that we observe switching of the IrMn$_3$ in both cases, i.e., while applying the writing pulse perpendicular or parallel to the reading current, is a further indication that the IrMn$_3$ pillar has a multidomain structure that can be manipulated by current pulses applied in either direction. Note that in this case (Fig. 2b), the peak-to-peak variation of the readout voltage ($V^{2-3}_{read}$) is approximately 20 µV, which is bigger than in the previous case (~ 8 µV in Fig. 2a). This suggests that the current-induced motion of domain walls is more efficient in the case of horizontal currents ($I^{1-4}_{write}$) compared to vertical currents ($I^{2-6}_{write}$). Given that the current density flowing in the Pt layer is the same in both cases, we attribute this to an increase in the thermally activated depinning and motion of domain walls due to Joule heating, given the larger resistance of the longer horizontal device arm (~ 970 Ω) compared to the shorter vertical one (~ 480 Ω). The difference of the resistance in the two cases is in good correspondence with the respective distance between the measurement electrodes, which is 70 µm for the longer arm and 30 µm for the shorter one, respectively. Thermal activation effects and the depinning of domain walls are discussed in more detail in the next section.

Finally, we studied the switching behavior under the same conditions (20 mA write current and 1 ms pulse width), using a 90˚ switching configuration. In this case, the write currents are applied perpendicular to each other either in the left cross ($I^{1-4}_{write}$ and $I^{6-2}_{write}$, labeled IrMn$_3$ in Fig. 3 with data points indicated in squares) or in the right cross of the device ($I^{1-4}_{write}$ and $I^{5-3}_{write}$, labeled Pt in Fig. 3 with data points indicated in circles). In each case, the voltages that are read out after horizontal write currents ($I^{1-4}_{write}$, which always passes through the IrMn$_3$ cross) are indicated in red, while the voltages that are read out after vertical write currents ($I^{6-2}_{write,}$, which passes through the IrMn$_3$ arm, and $I^{5-3}_{write}$, which passes through the Pt-only arm) are indicated in blue and dark blue, respectively. As expected, all cases indicate a clear switching signal, except the case of dark blue circles, which is the only case where the write current ($I^{5-3}_{write}$) does not pass through the IrMn$_3$ pillar.

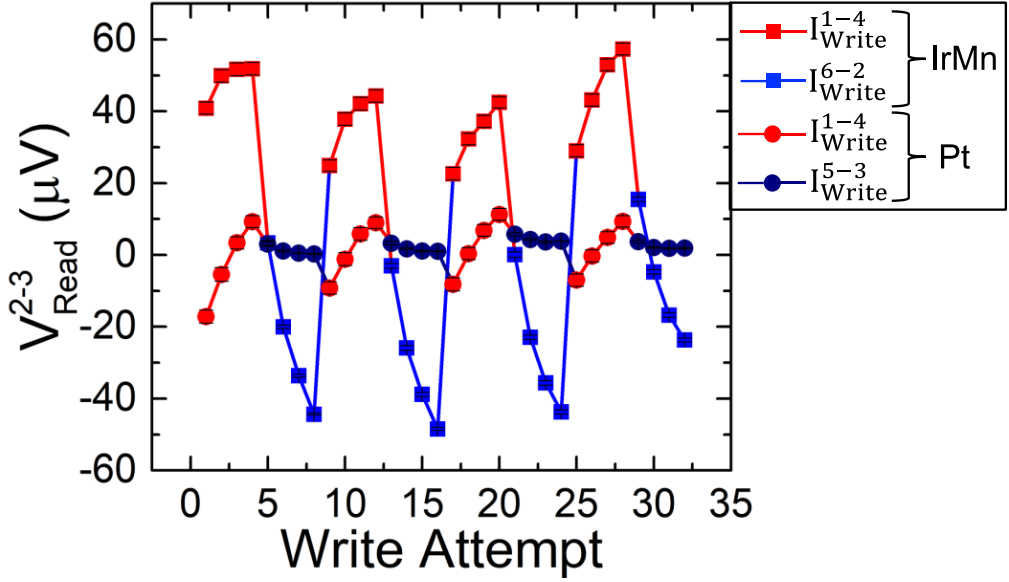

**Fig. 3. Measured differential voltage ($V^{2-3}_{Read}$) for the 90° switching configuration with 20 mA write pulses.** Square data points indicate the case where the write current pulses are applied in the cross containing the IrMn$_3$ pillar ($I^{1-4}_{write}$, red squares, and $I^{6-2}_{write}$, blue squares), and circle data points correspond to the case when the pulse is applied to the other cross ($I^{1-4}_{write}$, red circles, and $I^{5-3}_{write}$, dark blue circles). Note that one of the writing currents in the latter case ($I^{1-4}_{write}$) also passes through the cross with the IrMn$_3$ pillar. The differential voltage, $V^{2-3}_{Read}$, only shows switching in the cases where the current pulse passes through the cross containing the IrMn$_3$ pillar, i.e. all cases except $I^{5-3}_{write}$. Raw data for this figure (prior to subtraction of the background slope) can be found in Supplementary Note 3.

Specifically, when the 90° writing scheme is used in the arm containing the IrMn$_3$ pillar, we observe a clear modification of the output differential voltage, as in the previous cases of 180° switching (square points in Fig. 3). However, when performing the same experiment on the right arm, only the horizontal write pulse between electrodes 1 and 4 (which also passes through the IrMn$_3$ pillar) modifies the output voltage significantly, while the vertical write pulse does not. This result provides further clear evidence of the manipulation of the magnetic state of the IrMn$_3$ pillar by SOT. Note that a similar change in the output voltage is detected when the current is applied between electrodes 1 and 4 (red and dark red points) in both cases. However, note that blue and dark blue data points in the two graphs correspond to currents along different short arms of the device, i.e. the arm containing the IrMn$_3$ pillar ($I^{2-6}_{write}$) or the Pt arm without the IrMn$_3$ pillar ($I^{3-5}_{write}$). As a result, the initial state of the IrMn$_3$ Néel vector (before each application of $I^{1-4}_{write}$) is different in the two cases. Therefore, the absolute values of the readout voltage (red and dark red points) in the two cases are not the same. It is noteworthy that the average voltage difference for this case (~ 80 µV) is bigger than the 180° experiments reported above.

To further rule out any non-magnetic switching effects from the Pt layer, we repeated all of the above-mentioned measurements in a double-cross structure made only of a single 5 nm Pt layer (without any $IrMn_3$ pillar on either cross), using the same write current amplitude (20 mA) and pulse width (1 ms) as in the three previous switching experiments. The results are shown in Supplementary Note 4 and clearly indicate no switching in any of the three configurations. This confirms that the origin of switching, for 20 mA write currents in our devices (corresponding to a current density of 40 MA/cm$^2$), is the manipulation of the magnetic order in the antiferromagnetic $IrMn_3$ pillar.

Next, we performed switching experiments in the same Pt-only double-cross device using a 30 mA write current amplitude. The results are shown in Fig. 4. In this case (corresponding to a higher current density of 60 MA/cm$^2$ in the Pt layer), the measurements show a sawtooth-like signal from the Pt layer itself, which is not of magnetic origin.

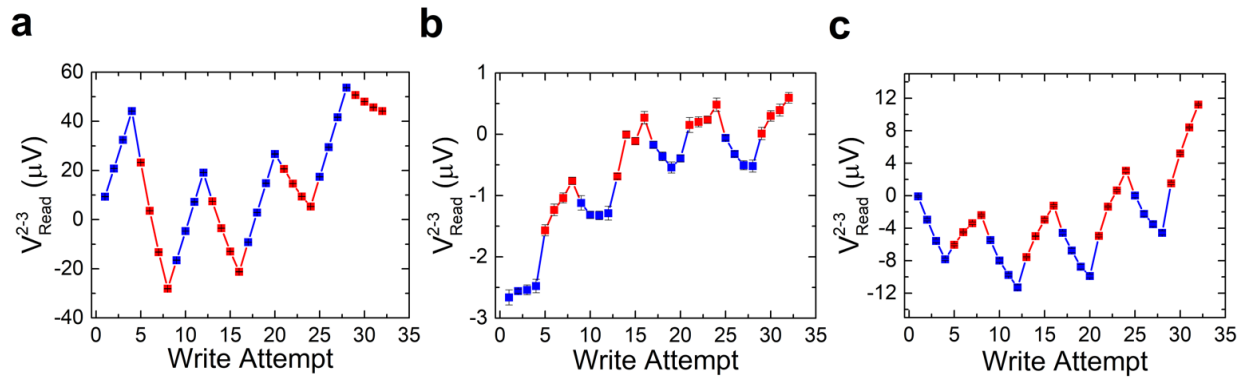

**Fig. 4. Measured differential voltage ($V^{2-3}_{Read}$) for a Pt control device with 30 mA write pulses. a** 90° switching configuration when the writing current is applied from electrode 1 to 4 ($I^{1-4}_{write}$, red squares) and from electrode 6 to 2 ($I^{6-2}_{write}$, blue squares), **b** 180° switching when the current pulse is applied from electrode 2 to 6 ($I^{2-6}_{write}$, red squares) and its reverse ($I^{6-2}_{write}$, blue squares), and **c** 180° switching between electrodes 1 and 4 (where $I^{1-4}_{write}$ corresponds to red squares and $I^{4-1}_{write}$ corresponds to blue squares). In all three cases, we observe a sawtooth-like switching of non-magnetic origin at this high writing current, which corresponds to a current density of 60 MA/cm$^2$ in the Pt layer.

Hence, to obtain a measurable switching signal in Pt-only double-cross devices, we need to apply a current amplitude of 30 mA (60 MA/cm$^2$), whereas the IrMn$_3$ switching in all the measured devices occurs for lower current densities. This is in agreement with the results of Figs. 2 and 3 (all at 20 mA), where no Pt switching was observed, and further indicates that the observed switching in Figs. 2 and 3 is due to AFM order manipulation, and hence has a different physical origin than that in Fig. 4, which is not magnetic.

*Micromagnetic simulations*

To better understand the experimental results, and in particular the role of the domain structure of the IrMn$_3$ pillar, we performed micromagnetic simulations. The micromagnetic framework that was used is based on two strongly coupled Landau-Lifshitz-Gilbert (LLG) equations (see Supplementary Note 5 for a detailed description of the model). In the simulations, the IrMn$_3$ thin film has a cubic anisotropy [50, 51] as shown in Fig. 5a. This figure plots $1-\varepsilon_{anis}$ rescaled to the range from 0 to 1, where $\varepsilon_{anis}$ is the anisotropy energy density. In this anisotropy configuration, there are 8 possible domain types oriented along the diagonals of the cubic cell, which are labeled in Fig. 5a as 1, 2, 3, 4, and their opposite directions. Note that the red color in Fig. 5a indicates a more stable state, favored by the anisotropy. From the x-ray diffraction measurements (see Supplementary Note 1), it is evident that one of the $(111)$ directions is primarily oriented along the out-of-plane direction, which we define as the *z*-direction. If we further assign the horizontal arm of the device to be the *x*-axis and the vertical one to be the *y*-axis (see Fig. 1 for the coordinate system), we can transform the crystallographic reference system to the new *xyz* reference system by the three Euler angles $\phi = \pi/4$, $\theta = \sin^{-1}(2/\sqrt{6})$, and $\psi = \pi/4$. Among these eight types of domains, six different domain walls (DWs) can in principle be present. Figs. 5b-g show one sublattice magnetization for these DWs as seen from the top view (+*z*), and from an in-plane view (-*y*). It should be noted that although the DW between 1 and -4 domain types is a possible rotation, the corresponding one between 1 and 4 domains is not. In the latter case, in fact, the most direct rotation from one domain to another requires the magnetization to be oriented along a high-energy direction $(001)$. Hence, a more favorable transformation requires the existence of an intermediate 2 or 3 domain state.

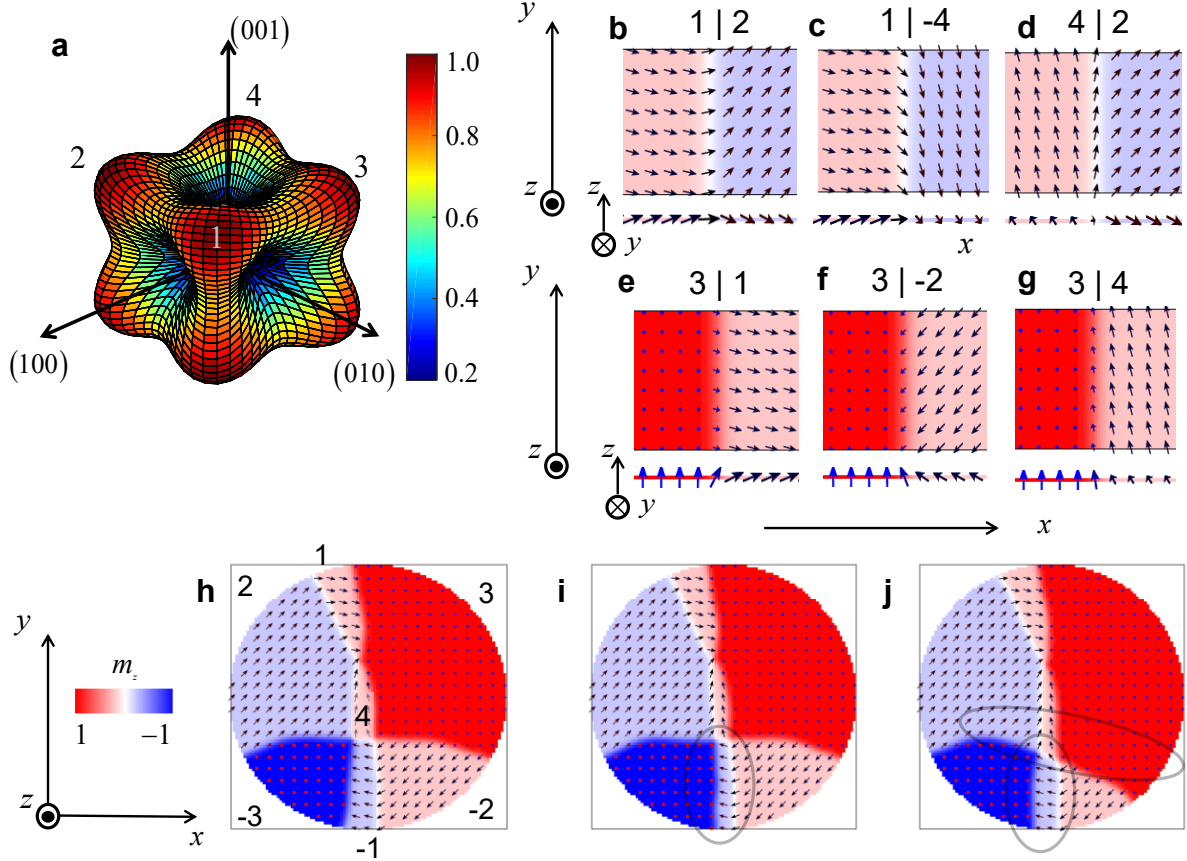

**Fig. 5. Cubic anisotropy and domains projected to the laboratory reference system. a** Cubic anisotropy energy density (with red colors indicating a more stable direction). The coordinate system refers to the crystallographic directions. **b-g** Snapshots of the first sublattice magnetization depicting the six possible domain walls in the laboratory reference system $x$ (horizontal arm in the real device), $y$ (vertical arm in the real device), and $z$ (out-of-plane direction). The snapshots correspond to the out-of-plane direction (+$z$) and an in-plane view (-$y$), as indicated by the arrows. **h** Multidomain state in a 500 nm diameter pillar relaxed from an initial random state. **i** The multidomain state after the application of a current pulse along the horizontal arm with an amplitude of 50 MA/cm$^2$ and width of 0.15 ns. **j** Multidomain state after the application of a current pulse with the same characteristics but opposite polarization.

To study the dynamical properties of the DWs, we first performed a set of micromagnetic simulations to calculate the behavior of each of the six DWs (Figs. 5b-g) by itself, in the presence of an electric current [52]. The DW velocity increases linearly as a function of the current, with mobility that depends on the DW type because of the different relative orientation between the DW magnetization and the spin polarization $p$ (perpendicular to the applied current). Table 1 summarizes a comparison of the DW velocities for a current density of 50 MA/cm² applied along the horizontal (row 1) and vertical (row 2) arms of the device.

| | DW velocity (m/s) | | | | | |
|---|---|---|---|---|---|---|
| | 3\|-2 | 3\|1 | 3\|4 | 1\|2 | 1\|-4 | 4\|2 |
| Horizontal ($p \parallel y$) | -359.5 | +488.0 | -131.2 | +282.7 | +200.2 | +81.3 |
| Vertical ($p \parallel x$) | +357.0 | +131.4 | -487.5 | -75.4 | +206.5 | -275.3 |

**Table 1.** Velocity of the six stable DW configurations when a current density of 50 MA/cm² is applied along the vertical and horizontal arms of the device. The velocities are expressed in m/s.

The torque is maximized when the spin polarization vector $p$ and the sublattice magnetizations inside the DW are perpendicular to each other. In fact, DWs involving a type 3 domain can reach a larger velocity because the DW magnetization has a significant out-of-plane component. In addition, DWs with an in-plane magnetization aligned along an intermediate orientation between $x$ and $y$ exhibit similar DW velocity for both horizontal and vertical current pulses.

To qualitatively describe the experimental results, we included in the micromagnetic simulations pinning forces in the form of anisotropy grains having a spatial distribution computed with a Voronoi tessellation algorithm. For each grain, the anisotropy constant was computed considering a Gaussian distribution with mean $|K_c| = 6.2 \times 10^6$ erg/cm³ and standard deviation $\sigma$. The resulting pinning potential introduces a threshold current to move the DWs, which depends on the standard deviation of the anisotropy distribution. In order to obtain the experimental threshold current density of about 36 MA/cm² (see next section), a standard deviation of $\sigma \approx 0.02$ was required in the simulations (see Supplementary Note 6 for more details).

Fig. 5h shows an example of a ground state of the IrMn₃ pillar (having a diameter of 500 nm) computed from micromagnetic simulations as a relaxation process starting from a random state. A standard deviation of $\sigma = 0.15$ was chosen in this case to speed up the simulations (see Methods and Supplementary Notes 5 and 6). We observe five of the six possible DWs in this figure (the 1|-4 case is missing). It can be also noted that the transition between the domain types 2 and 3 is mediated via a domain type 1 or 4. Fig. 5i shows the response of this ground state to a current pulse of 120 MA/cm$^2$ amplitude and 0.15 ns width, while Fig. 5j shows the state after a subsequent pulse of the same amplitude and opposite polarity. Current-induced DW motion is clearly observed in the regions enclosed by the grey ellipse in Fig. 5i. Nonetheless, it is worth noting that most of the DWs remain pinned in other parts of the pillar, and hence the change of the overall (average) Néel vector orientation in the pillar can be attributed to only a small portion of its area. In the case of the current pulse with reverse polarity, one can observe that the DW that had been previously displaced comes back to its prior position, while at the same time other DWs also move between pinning sites in the pillar, as marked by the grey ellipses in Fig. 5j. This partial movement of DWs between pinning centers in the pillar, in response to currents of different polarities and directions, can qualitatively explain the observed gradual (sawtooth-like) change of the readout voltage in Figs. 2 and 3, which indicates a gradual change of the average Néel vector in response to current, without reaching a saturated single-domain AFM state.

**Discussion**

Two observations from the above experiments are worth noting at this point. First, we observed a sawtooth switching behavior (instead of plateaus) in the IrMn$_3$ pillar. In a number of recent experiments [32, 39], sawtooth signals had been shown to be artifacts from heavy metals in AFM/HM structures at high current densities. The present measurements, however, show that under the same conditions (20 mA write current), Pt alone does not exhibit any variation of the output voltage. Hence, the presence of a sawtooth signal by itself is not a reliable indicator of a non-magnetic artifact in the switching experiments. In the present experiment, we interpret the presence of the sawtooth signals (where the output voltage depends on the number of applied write pulses in the same direction) to indicate the thermally-assisted motion of domain walls between pinning centers. This interpretation is consistent with the micromagnetic simulations shown in the previous section, as well as with the second key observation that can be made from Figs. 2 and 3, where one sees larger variations of the differential output voltage when the longer, more resistive horizontal arm is used to apply the write current.

To further investigate this hypothesis and measure the minimum current required for switching in this IrMn$_3$/Pt device, we repeated the same experiments at different current amplitudes in the range from 2 to 20 mA. The procedure consisted of applying three write attempts (as defined in the previous section), and measuring the differential voltage after each attempt, using a reading current of 100 µA. After this, we reversed the direction of the writing pulse and repeated the same procedure at the same amplitude, but in the opposite direction. These steps were repeated while increasing the write current amplitude in 1 mA steps and keeping the pulse width constant at 1 ms. The results following this protocol for the 180° switching scheme in the vertical (terminals 2-6) and horizontal (terminals 1-4) write current directions are shown in Fig. 6.

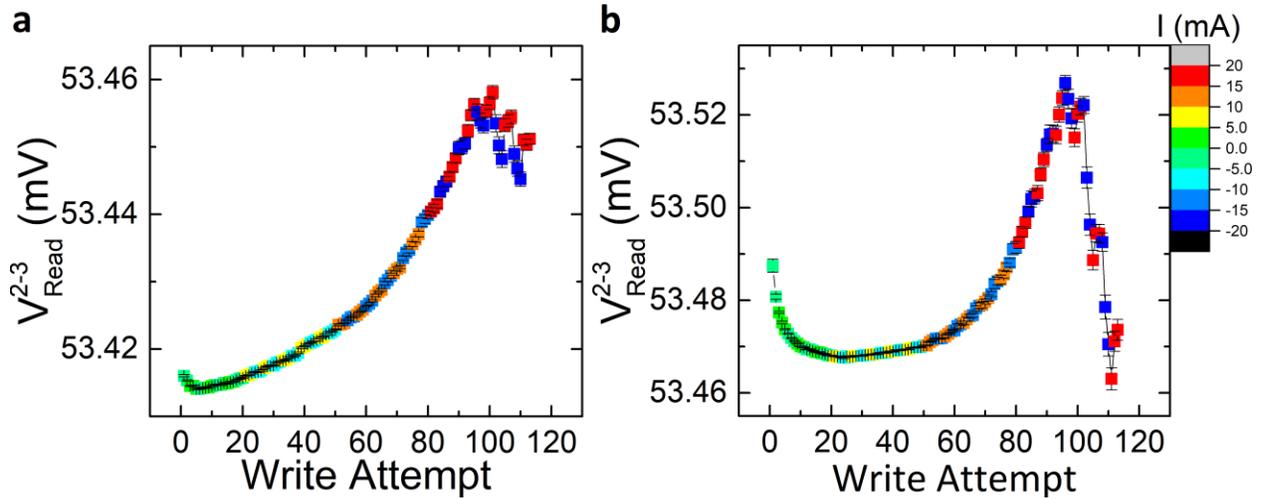

**Fig. 6. 180° switching in IrMn$_3$/Pt pillar, measured as a function of the write current amplitude.** The color map corresponds to the amplitude and direction of the writing current applied before the differential voltage measurement. The graphs correspond to cases **a** when the current pulse is applied between electrodes 2 and 6, $I^{2-6}_{write}$, and **b** when the writing current is applied between electrodes 1 and 4, $I^{1-4}_{write}$. For both cases, the device resistance increases due to Joule heating as we apply more write attempts, and for current pulses above 18 mA the output voltage shows a switching behavior, indicating a threshold current density of ~ 36 MA/cm$^2$ for switching in this structure.

For both cases, the total resistance of the device shows a non-linear increase as a function of the amplitude of the writing pulse, followed by a resistance peak. A similar trend in the resistance has been previously measured in MnTe thin films as a function of temperature [53, 54]. This suggests that the measured resistance background in Fig. 6 is associated with the Joule heating of the device by the applied current, and its effect on the AFM state, namely the spin fluctuations of the Mn atoms which are bigger when the device temperature is closer to the Néel temperature. It is worth noting that similar behavior has also been measured in the tunnel anisotropic magnetoresistance of Ir$_{0.2}$Mn$_{0.8}$/MgO/Ta magnetic tunnel junctions [55], as well as in the Seebeck coefficient of thin IrMn$_3$ films [56]. In the latter work, the Néel temperature of IrMn$_3$ is measured to be ~ 350 K for a 4 nm thin film, while the Néel temperature for bulk IrMn$_3$ is around 700 K.

Based on this, while we did not perform systematic measurements of the Néel temperature in our samples, we can expect our films to have a Néel temperature within the above range. Considering that the measured temperature increase due to a single 20 mA writing current pulse in our samples is ~96 K as shown in Supplementary Note 6, it is reasonable to ascribe this resistance trend to the temperature increase of the IrMn$_3$ pillar.

Note that the increase of the differential voltage is also larger when the write pulse is applied parallel to the reading current direction, i.e. terminals 1-4, which has a larger electrical resistance compared to the vertical arms, thus also generating more Joule heating. However, for both cases ($I^{1-4}_{write}$ and $I^{2-6}_{write}$), the threshold current required for switching in the IrMn$_3$ pillar is approximately the same, i.e. ~ 18 mA, corresponding to a threshold current density of ~ 36 MA/cm$^2$.

In summary, the experiments reported here provide clear evidence of the electrical switching of Néel order in an IrMn$_3$ pillar by spin-orbit torques from an adjacent Pt layer. The proposed differential measurement approach can be applied to a variety of AFM/HM material systems in a straightforward manner, including those with insulating AFM films. We expect that this will allow for reliable electrical measurement of the resistive signatures of AFM switching induced by SOT, which will be important not only for the fundamental understanding of current-induced AFM domain dynamics, but also as an electrical readout method for practical applications in memory and computing devices.

Finally, it is worth mentioning that the range of current densities where AFM switching can be observed without artifacts from the HM layer depends on the threshold switching current density of the AFM. Reducing the latter will allow for the operation of AFM/HM memory devices at even lower currents than those presented here, where the possibility of resistive signals due to the HM layer will be further reduced. One possible pathway to do so could be the further reduction of the IrMn$_3$ thickness compared to the 10 nm in this work. In particular, it is worth noting that Néel temperature reduction as a function of the thickness of IrMn$_3$ has been previously observed. Bulk IrMn$_3$ has a Néel temperature above 700 K [57], which can be reduced to room temperature for thickness values below 3 nm [55, 56]. The reduction of the Néel temperature (while still keeping it above room temperature by selecting an appropriate IrMn$_3$ thickness) will be accompanied by a reduction of the exchange constant, which in turn affects the threshold current. Thus, reducing the thickness of IrMn$_3$ may result in a reduction of the threshold current density to switch the AFM state, providing an even wider range of current densities where artifact-free AFM switching can be electrically observed.

## Methods

### *Materials and sample fabrication*

The materials used in this study were deposited on thermally oxidized silicon substrates, using sputter deposition under an Ar pressure of 2.5 mTorr at room temperature, in a physical vapor deposition system with base pressure less than $1 \times 10^{-8}$ Torr. The IrMn$_3$ films were deposited by co-puttering from Ir and Mn targets at 10 W and 40 W, respectively. All samples were capped with a 2.5 nm MgO protecting layer. The Pt(5)/IrMn$_3$(10) structure (thicknesses in nanometers), was patterned into the six-terminal devices shown in Fig. 1, using photolithography and dry etching techniques. The IrMn$_3$ was patterned on top of the Pt double cross as a micro-pillar with nominal diameters of 4 and 6 µm. Finally, after etching of the heavy-metal layer and the pillar, Cr(5)/Au(80) electrodes were grown by electron beam evaporation to form the electrical contacts of the device.

### *Magnetic characterization*

To confirm the antiferromagnetic character of the sputtered IrMn$_3$ films, a test structure of Ta(5)/Pt(8)/Co(0.8)/Pt(1)/IrMn$_3$(7)/MgO/Ru (thickness in nanometers) was magnetically characterized using a vibrating sample magnetometer (VSM) measurement system from Lake Shore. The samples were characterized in an out-of-plane field range between -500 and 500 Oe. The results show an exchange bias of ~ 170 Oe, confirming the AFM character of the IrMn$_3$ layer used in the experiments. Details of the magnetic characterization are shown in Supplementary Note 2.

### *Electrical measurements*

Electrical current pulses were applied in the six-terminal devices using two 6221 Keithley current sources for the longitudinal ($I^{1-4}_{write}$) and vertical pulse directions ($I^{2-6}_{write}$ and $I^{3-5}_{write}$). The reading current was always set to 100 µA between electrodes 1 and 4, as shown in Fig. 1. The differential voltage was measured between electrodes 2 and 3 using a Keithley 2182A nano-voltmeter while applying the reading current, using the delta mode option of the nano-voltmeter.

*Micromagnetic simulations*

In our micromagnetic model, we consider IrMn3 to have a face-centered cubic unit cell with lattice constant $a = 0.4$ nm[51], where only the Mn atoms are magnetic [58]. Therefore, the magnetic unit cell is a bcc cell with a lattice constant $a_m = a\sqrt{2}/2 \approx 0.28$ nm[51]. Even though IrMn3 is a noncollinear antiferromagnet with three sublattices, at a mesoscopic scale it can be described by means of two strongly coupled LLG-Slonczewski equations. In this framework, the inhomogeneous intra-lattice exchange constant is $A_{11} = 4.55 \times 10^{-7}$ erg/cm, and both the homogeneous and inhomogeneous inter-lattice exchange constants are $A_0 = A_{12} = -2.286 \times 10^{-6}$ erg/cm [50]. We considered the damping parameter to be $\alpha = 0.1$ [50], while the spin Hall angle is $\theta_{SH} = 0.1$ [59]. The saturation magnetization of Mn was considered to be $M_S = 153.75$ emu/cm$^3$. Finally, the anisotropy is known to be cubic, with anisotropy constant $K_c = -6.2 \times 10^6$ erg/cm$^3$[50].

**Data Availability**

The data that support the plots within this paper and other findings of this study are available from the corresponding authors upon reasonable request.

**Acknowledgements**

This work was supported by a grant from the National Science Foundation, Division of Electrical, Communications and Cyber Systems (NSF ECCS-1853879). This work was also supported by the National Science Foundation Materials Research Science and Engineering Center at Northwestern University (NSF DMR-1720319) and made use of its Shared Facilities at the Northwestern University Materials Research Center. One of the magnetic probe stations used in this research was supported by an Office of Naval Research DURIP grant (ONR N00014-19-1-2297). This work also utilized the Northwestern University Micro/Nano Fabrication Facility (NUFAB), which is partially supported by the Soft and Hybrid Nanotechnology Experimental (SHyNE) Resource (NSF ECCS-1542205), the Materials Research Science and Engineering Center (NSF DMR-1720139), the State of Illinois, and Northwestern University. For part of the sample fabrication, use of the Center for Nanoscale Materials, an Office of Science user facility, was supported by the US Department of Energy, Office of Science, Office of Basic Energy Sciences, under contract no. DE-AC02-06CH11357. G.F., L.S.-T. and F.G. also acknowledge support from PETASPIN.

**Author contributions**

G.F., V.L.-D., and P.K.A. conceived the idea and designed the devices. S.A. and V.L.-D. deposited the materials. S.A. and J.S. fabricated the devices. V.L.-D., S.A., and J.S. performed the measurements. L.S.-T. implemented the routines to include the IrMn$_3$ anisotropy in the PETASPIN micromagnetic solver. L.S.-T. and F.G. performed the micromagnetic simulations. G.F. coordinated the simulation activities. C.W., X.Y. and M.G. performed the temperature increase characterization for the current-induced switching measurements. V.K.S. and M.C.H. performed the thermal treatment on the double-pillar samples and some of the related electrical measurements. P.K.A., G.F., V.L.-D. and S.A. wrote the manuscript with contributions from the other authors. All authors discussed the results, contributed to the data analysis and commented on the manuscript. S.A. and V.L.-D. contributed equally to this work. The study was performed under the supervision of P.K.A.

**Competing interests**

The authors declare no competing interests.